\documentclass[twocolumn,showpacs,preprintnumbers,amsmath,amssymb,10pt,fleqn]{revtex4}

\usepackage{euscript}

\begin{document}

\title{Observational Limits on Gauss-Bonnet and Randall-Sundrum Gravities}

\author{Stanislav O. Alexeyev$^1$, Kristina A. Rannu$^1$, Polina I. Dyadina$^2$, Boris N. Latosh$^3$,$^{4}$, and Slava G. Turyshev$^{5}$}

\affiliation{\vskip 3pt
$^1$Sternberg Astronomical Institute, Lomonosov Moscow State University, Universitetsky Prospekt, 13, Moscow 119991, Russia}

\affiliation{\vskip 3pt
$^2$Physics Department, Lomonosov Moscow State University, Vorobievy Gory, 1/2, Moscow 119991, Russia}

\affiliation{\vskip 3pt
$^3$Faculty of Natural and Engineering Science, Dubna International University, Universitetsaya Str., 19, Dubna, Moscow Region, Russia\\
$^4$Physics Department, Institute for Natural Sciences, Ural Federal University, Kuibyshev Str., 48, Yekaterinburg, Russia}

\affiliation{\vskip 3pt
$^5$Jet Propulsion Laboratory, California Institute of Technology, 4800 Oak Grove Drive, Pasadena, CA 91109-0899, USA}

\date{December 22, 2014}

\begin{abstract}

We discuss the possibilities of experimental search for new physics predicted by the Gauss-Bonnet and the Randall-Sundrum theories of gravity. The effective four-dimensional spherically-symmetrical solutions of these theories are analyzed. We consider these solutions in the weak-field limit and in the process of the primordial black holes evaporation. We show that the predictions of discussed models are the same as of General Relativity. So, current experiments are not applicable for such search therefore different methods of observation and higher accuracy are required.

\end{abstract}



\maketitle

\section{Introduction}

A set of multidimensional gravity models beginning from the Kaluza-Klein one \cite{KK} result from the attempts to construct an unified field theory. As we live in a space-time with four noncompact
dimensions any multidimensional theory needs an appropriate effective four-dimensional limit consistent with the predictions of general relativity (GR) and the results of observations and experiments.

String theory \cite{scherk} along with the loop quantum gravity \cite{loop} is a promising candidate for a quantum theory of gravity nowadays. Lovelock gravity \cite{lovelock} appeared to be a ghost-free four-dimensional low-energy effective limit of the string theory \cite{string, wheeler2}:
\begin{gather}\label{eq:01}
  L = \sqrt{- g} \left( R + \alpha_2 S^2 + \alpha_3 S^3 + \dots \right),
\end{gather}
where $S^n$ is Euler characteristic of the n-th order. The leading and the most studied among them is the second order curvature correction given by a Gauss-Bonnet term $S^2 = S_{GB} = R_{ijkl} R^{ijkl} - 4 R_{ij}R^{ij} + R^2$. The effective four-dimensional limit of the string theory also includes the scalar field that is the projection of the $g_{10\ 10}$ component of the ten-dimensional string metric to the four-dimensional manifold. The Gauss-Bonnet term coupled to the scalar
(dilatonic) field \cite{alex1, alex2, ghs, mignemi, kanti} describes the influence of the compact extra dimensions on the four-dimensional space-time. Therefore the Gauss-Bonnet theory with the dilaton scalar field serves as the effective four-dimensional limit of the string theory.

Unlike the string theory the Randall-Sundrum model allows the only extra dimension to be large and even infinite \cite{rs1,  rs2}. This model considers four-dimensional branes with tension
embedded into a five-dimensional space-time (bulk) that is assumed to have an AdS$_5$ geometry. All the matter and the three fundamental interactions are localized on this brane except gravity which is allowed to propagate into the bulk along the extra noncompact dimension. Thus the Randall-Sundrum model contains the description of the four-dimensional space-time from the very beginning and therefore does not need any special theory serving as its effective four-dimensional limit. The Randall-Sundrum I (RSI) model includes two branes with different properties helping to solve the hierarchy problem \cite{rs1}. Moving the second brane to the infinity led to the 
Randall-Sundrum II (RSII) model with one brane \cite{rs2}. In this paper we deal with RSII only.

Black hole solution is a basic one for any theory of gravity. First of all, it describes the compact object that a very massive star at the end of its life cycle collapses into. It also features the curvature of the space-time produced by the presence of matter specific for the considered gravity model. Any extended theory of gravity should be consistent with the predictions of GR and the observational results therefore the existence of black holes and their properties are important indicators of the theory's viability.

The Gauss-Bonnet solution has been studied explicitly during the recent years \cite{ghs, mignemi, kanti, alex1}. On the contrary there are several different solutions for Randall-Sundrum model \cite{dadhich1, dadhich2, maartens1, rs_bh, kudoh}. It was argued that static black holes cannot exist in RSII with a radius much greater than the AdS length $\ell$ \cite{tanaka1, emparan1, emparan2} and that even very small RSII static black holes do not exist \cite{yoshino1, kleihaus1}. Therefore RSII solutions for large black holes that have been found independently by Figueras and Wiseman \cite{fw1, fw2} and Abdolrahimi, Catto\"en, Page and Yaghoobpour-Tari \cite{abd1} are
an important improvement of the Randall-Sundrum model interesting for further consideration. In their work \cite{abd1} Abdolrahimi, Catto\"en et al. compare the obtained black-hole solution to the one Figueras and Wiseman \cite{fw2} and show that these solutions agree closely. In this paper we use the ACPY solution \cite{abd1} as it contains the necessary details. The Figueras-Wiseman solution \cite{fw1} is considered separately.
 
The outline of this paper is as follows: in Section ``Weak field limit'' we discuss the weak field and slow motion approximation of the Gauss-Bonnet and Randall-Sundrum theories. Section ``Thermodynamics and PBHs'' is devoted to the analysis of thermodynamical properties of these models and their influence on the primordial black holes mass spectra. In Section ``Discussion and Conclusions'' we discuss the results obtained, offer conclusions, and outline the next steps.

\section{Weak field limit}

As a weak-field limit we consider the dynamical conditions in the solar system i.e. post-Newtonian approximation. The metric tensor $g_{\mu\nu}$ can be represented as the perturbation $h_{\mu\nu}$ around Minkowski space-time $\eta_{\mu\nu}$ \cite{will}:

\begin{gather}\label{eq:02}
  g_{\mu\nu} = \eta_{\mu\nu} + h_{\mu\nu}.
\end{gather}

In this paper we consider only spherically-symmetric solutions and therefore the static gravity field at the distance $r$ from its source. In the first post-Newtonian (PPN) order the correction to the
gravitational field $h_{\mu\nu}$ can be expressed by the series with respect to the negative powers of the radial coordinate $r$ up to the following order terms:
\begin{gather}\label{eq:03}
  h_{00} \sim {\cal O}(r^{-3}), \quad h_{0j} \sim {\cal O}(r^{-4}), \quad h_{ij} \sim  {\cal O}(r^{-2})
\end{gather}
and use geometric units $\hbar = c = G = 1$ with non-dimensional masses expressed in units of Plank mass.

The PPN limit is well tested by experiments \cite{will, turyshev}. The better the experimental accuracy becomes \cite{turyshev} the more opportunities to test small gravitational effects predicted by currently viable theories should appear. We use the expansion (\ref{eq:03}) to compare the magnitudes of the predicted effects in order to see if the specific effects of the considered solutions can be tested. As the PPN requires the weak-field limit approximation we apply our results to solar system where the post-Newtonian parameters (PPN-parameters) are measured with high precision \cite{will2}. Our results are inapplicable to the strong field limit.

\subsection{Gauss-Bonnet gravity}

We begin our consideration by exploring the weak field limit of the Gauss-Bonnet theory (Here and further during solving the equations we use the dimensionless Planckian units, only at the step of numerical estimation values we jump to the usual ones)
\begin{gather}\label{eq:04}
  \begin{split}
    S &= \cfrac{1}{16\pi} \int d^4 x \sqrt{-g} \Big[ - R + 2 \partial_{\mu} \phi \partial^\mu \phi + \\
    & \hskip 87pt + \lambda e^{-2\phi} S_{GB} + \dots \Big],
  \end{split}
\end{gather}
where $\phi$ is the potential of the dilatonic field, $S_{GB}$ is the Gauss-Bonnet term and $\lambda$ is the string coupling constant. For this purpose we construct a post-Newtonian parameterization of the static asymptotically flat spherically-symmetric Gauss-Bonnet solution
\begin{gather}
  ds^2 = \Delta dt^2 - \cfrac{\sigma^2}{\Delta} \ dr^2 - r^2 \Big(
  d\theta^2 + \sin^2\theta d\varphi^2\Big), \label{eq:05} \\
  \begin{split}
    \Delta &= 1 - \cfrac{2M}{r} + {\cal O} \left( r^{-2} \right),
    \quad \sigma = 1 + {\cal O} \left( r^{-2} \right), \\
    \phi &= \phi_{\infty} + \cfrac{D}{r} + {\cal O}\left( r^{-2} \right), 
  \end{split}\label{eq:06}
\end{gather}
where $(t, r, \theta, \varphi)$ are usual spherical coordinates and functions $\Delta$ and $\sigma$ depend on the radial coordinate $r$ only, $M$ is the Arnowitt-Deser-Misner (ADM) mass, $D$ is dilatonic charge i. e. the effective charge of the scalar field source and $\phi_{\infty}$ is the asymptotic value of dilatonic potential \cite{alex1, alex2}. As it was argued in \cite{mignemi} $D \propto 1/M$.

We put metric (\ref{eq:05}) with the expansions (\ref{eq:06}) to the field equations written down in the most computationally convenient form \cite{sotiriou}:
\begin{gather}\label{eq:07}
  G_{\mu\nu} = 8 \pi \left(T^m_{\mu\nu} + T^{\phi}_{\mu\nu} + T^{\rm GB}_{\mu\nu} \right),
\end{gather}
where $T^m_{\mu\nu}$ is a matter stress-energy tensor while $T^{\phi}_{\mu\nu}$ and $T^{\rm GB}_{\mu\nu}$ describe the scalar field and the Gauss-Bonnet term presence:
\begin{gather*}
  \begin{split}
    T^{\phi}_{\mu\nu} &= \cfrac{1}{8 \pi} \left( \partial_{\mu} \phi \ \partial_{\nu} \phi - \cfrac{1}{2} \
      g_{\mu\nu}  \partial^{\rho} \phi \ \partial_{\rho} \phi \right), \\
    T^{\rm GB}_{\mu\nu} &= \cfrac{1}{8 \pi} \ \Big[  (\nabla_{\mu}
    \nabla_{\nu}  -  g_{\mu\nu} \Box ) (e^{-2\phi}R) \\
    &+ \ 2 \ \Big(\Box \delta^\sigma_\mu \delta^\sigma_\nu + \
    g_{\mu\nu} \nabla^{\rho} \nabla^{\sigma}- \nabla^{\rho}
    \nabla_{(\mu}\delta^\sigma_{\nu)} \Big) (e^{-2\phi} \ R_{\rho\sigma}) \\
    &- \ 2 \nabla^{\rho} \nabla^{\sigma} (e^{-2\phi} \ R_{\mu\rho\nu\sigma}) \Big].
  \end{split}
\end{gather*}
Using the standard computational techniques \cite{will} the leading order for the nontrivial correction to the Gauss-Bonnet metric tensor is 
\begin{gather}\label{eq:08}
  \delta h_{00}^{\rm GB} = 8 \ \cfrac{DM}{r^4} + {\cal
    O}(r^{-5}).
\end{gather}

Comparing this result with (\ref{eq:03}) we see that the correction term (\ref{eq:08}) lies beyond the PPN order that should be proportional to $1/r^2$. Thus the parameters of the Gauss-Bonnet model
cannot be constrained by the solar system tests. This result is consistent with conclusions from \cite{sotiriou} where the cosmological limit of the discussed model was studied.

\subsection{Randall-Sundrum gravity}

The black hole solution of the Randall-Sundrum model was constructed by Figueras and Wiseman \cite{fw1} using an associated five-dimensional anti-de Sitter space (AdS$_5$) and dS$_5$-CFT$_4$
correspondence \cite{haro}. The Figueras-Wiseman solution describes a static black hole with radius up to $\sim 20 \ell$ and reproduces four-dimensional GR on the brane in the low curvature and the low energy limit. We intend to use the fact that Schwarchild metric can be used not only as a black hole one but also as a description of gravitational (stellar) system far from the central body. For example, for Solar system (taking into account all the limitations and corrections).

The five-dimensional metric can be written near the conformal boundary
$z = 0$ as
\begin{gather}\label{eq:09}
  ds^2 = \cfrac{l^2}{z^2} \left[ d z^2 + \tilde{g}_{\mu\nu} (z,x) \ dx^{\mu} dx^{\nu} \right],
\end{gather}
where $z$ is a coordinate of the brane along the extra dimension and $\tilde{g}_{\mu\nu} (z,x)$ is the metric on the brane determined by the Fefferman-Graham expansion \cite{haro}. The corresponding effective four-dimensional field equations \cite{fw1} are:
\begin{gather}\label{eq:10}
  \begin{split}
    G_{\mu\nu} &= 8 \pi G_4 T_{\mu\nu}^{brane} + \epsilon^2 \Big\{
    16 \pi G_4 \langle T_{\mu\nu}^{CFT}[g] \rangle + \\
    &+ a_{\mu\nu}[g] + \log \epsilon \ b_{\mu\nu}[g] \Big\} +  O (\epsilon^4 \log \epsilon),
  \end{split}
\end{gather}
where $G_4$ is the usual four-dimensional gravitational constant, $T_{\mu\nu}^{brane}$ is the stress-energy tensor of the matter localized on the brane, tensors $\langle T_{\mu\nu}^{CFT}[g] \rangle$, $a_{\mu\nu}[g]$ and $b_{\mu\nu}[g]$ result from the extra dimension and depend on the metric tensor components and $\epsilon$ is a small perturbation parameter indicating the deviation of the brane position from the equilibrium state $z = 0$.

The additional term in the post-Newtonian expansion of the Figueras-Wiseman solution calculated in this paper is 
\begin{gather}\label{eq:11}
  \delta h_{00}^{\rm FW} = \cfrac{121}{27} \ \cfrac{\epsilon^2}{\ell^2} \ \cfrac{M^2}{r^2}.
\end{gather}
The obtained value (\ref{eq:11}) lies within the PPN limit (\ref{eq:03}) and points at a potentially observable effect. In Randall-Sundrum model gravity is allowed propagate into the bulk along
the extra dimension therefore the effect described by (\ref{eq:11}) most likely leads to the negative nonlinearity in gravitational superposition. In other words the resulting gravitational field
produced by two or more massive objects can be less than the direct vector sum of their contributions. The parameterized post-Newtonian (PPN) parameter $\beta$ is responsible for such an effect \cite{will, turyshev}. Therefore the result (\ref{eq:11}) should be expressed as follows:
\begin{gather}\label{eq:12}
  \beta = 1 - \cfrac{\epsilon^2}{\ell^2} \ \cfrac{121}{108} \ M^2,
\end{gather}
where $M$ is the mass of the massive central object. In the considered case it equals the solar mass. It also is expressed in Plank units of mass and therefore is dimensionless.

The constraint on the PPN parameter $\beta$ obtained from analysis of the lunar laser ranging data \cite{Williams-etal-2004} is $|\beta - 1|\leq 1.1 \times 10^{-4}$ \cite{will2}. The admitted region of the AdS length is limited by the results of the Newton's law test $\ell < 10^{-5}$~m \cite{kapner}. Therefore the upper limit on the value of $\epsilon$ is:
\begin{gather}\label{eq:13}
  \epsilon \leq 5.7 \times 10^{-47} \mbox{cm} \ll l_{Pl}.
\end{gather}
 
Originally the parameter $\epsilon$ was assumed to be negligibly small and the vanishing value found in (\ref{eq:13}) implies that in fact $\epsilon = 0$. Thus the Figueras-Wiseman four-dimensional black hole solution is not only is self-consistent but well consistent with the solar system constraints as well. Therefore this solution is indistinguishable form GR in the PPN limit after all.

The other recent Randall-Sundrum solution obtained by Abdolrahimi, Catto\"en, Page and Yaghoobpour-Tari (ACPY) \cite{abd1} is asymptotically conformal to the Schwarzschild metric and includes a negative five-dimensional cosmological constant $\Lambda_5$:
\begin{gather}\label{eq:14}
  \begin{split}
    ds^2 &= - \ u(r) dt^2 + \frac{v(r)}{u(r)} \ dr^2 + \left[ r^2 +
      \cfrac{F(r)}{- \Lambda_5} \right] d \Omega^2, \\
    u(r) &= 1 - {2M}/{r}, \\
    v(r) &= 1 + \left( \cfrac{r - 2M}{r - (3 M / 2)} \right) \left[ \cfrac{F(r)}{- \Lambda_5r} \right]', \\
    F(r) &= 1 - 1.1241 \left( \cfrac{2M}{r} \right) + 1.956 \left( \cfrac{2M}{r} \right)^2 - \\
    &\hskip 18pt - 9.961 \left( \cfrac{2M}{r} \right)^{3} + \ \dots \ + 2.900 \left( \cfrac{2M}{r} \right)^{11},
  \end{split}
\end{gather}
where $' \equiv d/dr$. The function $F(r)$ describes the perturbation caused by the bulk. The best fit for it was obtained in \cite{abd1}.

The field equations induced on the brane were derived by Sasaki, Shiromizu and Maeda \cite{maeda1}:
\begin{gather}\label{eq:15}
  G_{\mu\nu} = - \ \Lambda_4 g_{\mu\nu} + \cfrac{8 \pi}{M^2_{\rm
      Pl4}} \ T_{\mu\nu} + \cfrac{8 \pi}{M^3_{\rm Pl5}} \ S_{\mu\nu} - \EuScript{E}_{\mu\nu},
\end{gather}
where $\Lambda_4$ is usual four-dimensional cosmological constant, $g_{\mu\nu}$ is the metric on the brane, $T_{\mu\nu}$ is the stress-energy tensor of the matter localized on the brane,
$S_{\mu\nu}$ is the local quadratic stress-energy correction, $\EuScript{E}_{\mu\nu}$ is the four-dimensional projection of the five-dimensional Weyl tensor. $M_{\rm Pl4}$ is usual four-dimensional Planck mass and $M_{\rm Pl5}$ is the fundamental five-dimensional Planck mass.

The induced metric on the brane is flat, the bulk is an anti-de-Sitter Space-time as in the original Randall-Sundrum scenario \cite{rs2}, then $\EuScript{E}_{\mu\nu} = 0$ \cite{maartens1}. Therefore the correction term due to the contribution from ACPY topology (\ref{eq:14}) that follows from (\ref{eq:15}) has the form
\begin{gather}\label{eq:16}
  \delta h_{00}^{\rm AP} = \cfrac{\ell^2 M^2}{96} \ \cfrac{1}{r^4} + {\cal O}(r^{-5}).
\end{gather}

According to (\ref{eq:03}) the expansion term of the PPN-order should be proportional to $r^{-2}$. The correction (\ref{eq:16}) contains the next perturbation order which lies beyond PPN similarly to the Gauss-Bonnet case (\ref{eq:08}). Therefore the obtained contribution (\ref{eq:16}) cannot be observed in the solar system experiments as well. This conclusion on the Randall-Sundrum model predictions confirms the result for the Figueras-Wiseman and coincides with the Gauss-Bonnet case.

\section{Thermodynamics and PBHs}

It is conjectured that density fluctuations in the early Universe could have created black holes with arbitrarily small masses even to the Planck scale \cite{carr1}. These black holes are referred to as primordial black holes (PBHs) \cite{macgibbon_2013} and can be used to consider viable theories in cosmological conditions.

Hawking evaporation \cite{hawking, NovikovFrolov} is one of the most significant properties of a black hole and can be described by the mass-loss rate equation \cite{page}:
\begin{gather}\label{eq:17}
  - \cfrac{dM}{dt} = \cfrac{1}{256} \ \cfrac{k_B}{\pi^3 M^2},
\end{gather}
where $M$ is the mass of a black hole and $k_B$ is the Stefan-Boltzmann constant. Hawking evaporation is a quantum process forbidden in classical physic. An outgoing radiation has to cross a potential barrier of black hole horizon \cite{pw1} so a radiation surrounding a black hole is in a thermal equilibrium and can be described as a black body radiation. Therefore the black hole evaporation obeys the following law:
\begin{gather} \label{eq:18}
  - \cfrac{dM}{dt} = k_B S T^4,
\end{gather}
where $S$ is its surface area. We use this formula to estimate the lifetime of the black holes in the Gauss-Bonnet and the Randall-Sundrum models.

According to (\ref{eq:17}) the black holes with stellar masses evaporate very slowly and do not lose mass through this process noticeably. On the other hand PBHs with initial masses smaller than
\begin{gather}\label{eq:19}
  M_{\rm 0} \approx 5.0 \times 10^{14} \ \mbox{g}
\end{gather}
have already evaporated and can contribute to the extragalactic background radiation \cite{NovikovFrolov}. PBHs with initial mass greater than $M_0$ (\ref{eq:19}) should be evaporating till now \cite{carr2}. According to some models of black hole evaporation \cite{alex1, alex2, carr2} the last stages of this process can be accompanied by bursts of high-energy particles \cite{alex2} including gamma radiation with energy of the MeV-TeV range that occur at the distances about $z \le 9.4$ \cite{swift}. Such events should be rather rare and on the other hand the set of more simple explanations for most part of gamma ray bursts (GRB) exists. Nevertheless PBHs at the last stage of evaporation can serve as additional candidates for GRB progenitors and therefore the limit estimation for black hole evaporation rate can be obtained in such a way.

Different theories of gravity predict different black hole evaporation rate and therefore different initial masses of the PBHs that fully evaporate for the Universe lifetime. In this paper we compare the evaporation rates for the Gauss-Bonnet and the RSII black hole solutions. According to the GRB data and the precision of the Fermi LAT telescope the closest distance $d$ at which the telescope will be able to detect the evaporation of primordial black holes is \cite{macgibbon_2013}
\begin{gather}\label{eq:20}
  d \simeq 0.04 \left( \cfrac{\Omega}{\mbox{sr}} \right)^{-0.5} \left(
    \cfrac{E}{\mbox{GeV}} \right)^{0.7} \left( \cfrac{T}{\mbox{TeV}} \right)^{0.8} \mbox{pc},
\end{gather}
where $\Omega$ is the angular resolution of the telescope, $E$ -- energy range of the telescope, $T$ -- temperature of the black hole. Spending the same reverse procedure and using a telescope
detected gamma-ray bursts lead to the observable difference of the PBH initial mass on its final evaporation stage can vary from the GR predictions within the following limits:
\begin{gather}\label{eq:21}
  \cfrac{M_{\mbox{investigated theory}}}{M_{GR}} > 10^5. 
\end{gather}
Therefore we use this limit as the mass cutoff threshold in our calculations.

Using the method by Shankaranarayanan, Padmanabhan and Srinivasan \cite{pw2} it is possible to rewrite the expression for the Gauss-Bonnet black hole temperature and then use (\ref{eq:07}). In the astrophysical case the dilatonic charge $D \simeq 1/M$ \cite{mignemi}. Therefore the right part of (\ref{eq:18}) can be expanded in series as
\begin{gather}\label{eq:22}
   - \cfrac{dM}{dt} \simeq \cfrac{1}{256} \ \cfrac{k_B}{\pi^3 M^2} +
   \cfrac{1}{512} \ \cfrac{k_B}{\pi^3 M^6} + {\cal O} \left({M^{-10}}\right).
\end{gather}
The initial mass of the PBH that fully evaporates during the lifetime of the universe in this case is
\begin{gather}\label{eq:23}
  M_{\rm GB} = 8 \times 10^{14} \mbox{g}.
\end{gather}
The difference between the obtained value and the similar GR quantity (\ref{eq:19}) is smaller than the cutoff threshold set by (\ref{eq:21}). Thus the specific features of the Gauss-Bonnet evaporation rate are negligible at the current level of accuracy and the predictions of Gauss-Bonnet gravity for the Hawking evaporation are indistinguishable from those of the GR.

One of the first and most studied black hole solutions of Randall-Sundrum model was found by Dadhich, Maartens et al. \cite{dadhich1, maartens1}. They obtained an exact localized black
hole solution, which remarkably has the mathematical form of the Reissner-N{\"o}rdstrom solution, but without electric charge being present \cite{dadhich1}:
\begin{gather}\label{eq:24}
  - \ g_{tt} = g_{rr} = 1 - \cfrac{2M}{r} + \left( \cfrac{q}{M^2_{Pl5}} \right) \cfrac{1}{r^2},
\end{gather}
The Reissner-N{\"o}rdstrom-type correction to the Schwarzschild potential in (\ref{eq:24}) can be thought of as a dimensionless ``tidal charge'' parameter $q$, arising from the projection onto the
brane of free gravitational field effects in the bulk transmitted via the bulk Weyl tensor \cite{dadhich1}. The projected Weyl tensor, transmitting the tidal charge stresses from the bulk to the brane, is \cite{dadhich1}:
\begin{gather*}
  \EuScript{E}_{\mu\nu} = - \left( \cfrac{q}{M_{Pl5}^2} \right)
  \cfrac{1}{r^4} \left( u_{\mu} u_{\nu} - 2 r_{\mu} r_{\nu} + h_{\mu\nu} \right), 
\end{gather*}
where $h_{\mu\nu} = g_{\mu\nu} + u_{\mu} u_{\nu}$ projects orthogonal to 4-velocity field $u^{\mu}$ and $r_{\mu}$ is a unit radial vector.

The mass loss rate obtained similarly to the Gauss-Bonnet case equals to
\begin{gather}\label{eq:25}
  - \cfrac{dM}{dt} = \cfrac{1}{216} \ \cfrac{k_B}{\pi^3 M^2} + {\cal O} \left({M^{-6}}\right).
\end{gather}
The leading term in (\ref{eq:25}) cannot produce the needed $5$-order difference defined by the threshold parameter (\ref{eq:21}). The initial mass of the Dadhich-Rezania black hole that evaporates completely during the lifetime of the Universe proves this fact:
\begin{gather}\label{eq:26}
  M_{\rm DR} = 5.3 \times 10^{14} \mbox{g}.
\end{gather}
As the obtained difference is much less than the cutoff threshold (\ref{eq:21}) the ``tidal charge'' influence is vanishing and cannot have experimentally verifiable consequences.

The black hole evaporation for the ACPY solution discussed in the previous section also was considered in a similar manner. The evaporation rate of this solution has completely the same form as the original Hawking formula (\ref{eq:17}) up to $M^{-10}$ terms thus the value of the initial mass equals to that given by the GR: 
\begin{gather}\label{eq:27}
  M_{\rm AP} = 5.0 \times 10^{14} \mbox{g}.
\end{gather}
The results for Figueras-Wiseman solution are the same because of the form of the solution (\ref{eq:11}).

The obtained results (\ref{eq:22}, \ref{eq:23}, \ref{eq:25}-\ref{eq:27}) lead to a conclusion that the precision  of the currently existing GRB data is not sufficient to distinguish the GR, the Gauss-Bonnet and the Randall-Sundrum gravity from each other via the PBH consideration.

\section{Discussion and Conclusions}

In this paper we discussed the possibilities to test the theories extending GR in different ways by the example of the Gauss-Bonnet and the Randall-Sundrum models both in the weak field and the cosmological limits. For this purpose the post-Newtonian expansion and the black hole evaporation of these theories' solutions were considered.

As the Gauss-Bonnet term coupled with the scalar field does not influence the post-Newtonian limit (\ref{eq:08}) therefore the nontrivial scalar hair generated by it \cite{alex1, kanti} does not contribute into the required order of the spherically-symmetric solution's expansion. This result agrees with the previous conclusions by Sotiriou and Barausse \cite{sotiriou} who considered the cosmological solution of the action (\ref{eq:04}) and showed that the influence of the Gauss-Bonnet term is negligible at solar system scales. Combining these two results we can state that the leading term of Lovelock expansion (\ref{eq:01}) describing second-order curvature correction does not provide any visible deviation from GR predictions in the weak-field limit therefore such theory of gravity fully agrees with GR.

This conclusion is also valid for any model with higher-order curvature corrections having a proper Newtonian limit. As the Gauss-Bonnet term is the leading curvature correction of the Lovelock gravity its contribution into the post-Newtonian expansion of the metric also is the largest one. Taking into account the others Euler characteristics i.e. the next orders of curvature corrections isn't able to change the picture as their influence is even less and obviously lies far beyond the PPN limit. Thus the conclusions for the Gauss-Bonnet model can be generalized to the Lovelock gravity.

The theories with curvature power series are not the only method of geometrical extending of GR. In the generic case the Lanrangian can contain an arbitrary function of the Ricci scalar $R$. Such theories set up $f(R)$-gravity \cite{capo, fR} and Lovelock gravity is its particular case. Many $f(R)$-gravity models such as $\ln(R)$ or $1/R$ \cite{capo, lnr} were originally introduced as attempts to explain dark energy or dark matter. They don't have a proper PPN limit \cite{capo} and are inapplicable to solar system scale. Therefore our conclusions for Gauss-Bonnet theory in the weak-field limit are applicable for Lovelock gravity and $f(R)$-gravity of the Lovelock type.

The thermodynamical properties of the Gauss-Bonnet black hole solution were considered in details earlier \cite{mignemi, alex1,   alex2} however only the black holes of Planck scales were investigated. For the black holes with larger mass the influence of the Gauss-Bonnet term and the scalar field becomes negligibly small therefore the evaporation is predictably the same as in the GR case.

Since Randall and Sundrum proposed a theory of gravity with noncompact extra dimension \cite{rs1, rs2} several black hole solutions have been found \cite{rs_bh, dadhich1, fw1, fw2, abd1}. Consideration of the post-Newtonian expansion of the Figueras-Wiseman solution \cite{fw1} reveals such possible effect as a negative nonlinearity of gravitational superposition (\ref{eq:12}). It naturally results from the theory itself because gravity is allowed to propagate to the extra dimension in Randall-Sundrum model. However the breaking of gravitational superposition turns out to depend on a negligibly small parameter thus the predictions of the Figueras-Wiseman solution fully agrees with GR and the present observations. This effect may influence the strong field regime (close binary systems, black holes) as a consequence of curvature growth. So the next step could be the search of such features of the Randall-Sundrum model in the strong field limit. Fortunately this investigation is admissible as the large stable black hole solutions for RSII black holes have been found \cite{fw1, abd1}.

The consideration of the other black hole solution by Abdolrahimi, Page et al. \cite{abd1} shows that the terms describing the bulk influence (\ref{eq:16}) greatly exceed the limits of the post-Newtonian approximation. As a result both large Randall-Sundrum black holes solutions cannot be distinguished from the Schwarzschild metric at the solar system scales.

We have examined the evaporation rate for the Randall-Sundrum black holes as well. The results for one of the first solutions obtained by Dadhich, Maartens, Papadopoulos and Rezania \cite{dadhich1} and the latest one by Abdolrahimi, Page et al. \cite{abd1} are presented (\ref{eq:22}, \ref{eq:23}, \ref{eq:25}-\ref{eq:27}). The difference between the Dadhich-Rezania solution and the GR is negligibly small and the Page solution coincides with GR completely.

As is easy to see, many extended gravity models cannot be distinguished from GR and from each other neither at the solar system scales nor by the black holes thermodynamic properties. Therefore the coincidence of these extended theories with the GR serves a good argument in favor of their validity. However it does not mean that no difference can be found by other verification methods. Except the weak field and the cosmological tests a strong field approximation is widely used. It has such a verification laboratory as close binary systems first of all containing pulsars as one or even both its components. A great amount of data has been obtained from these observations and it obviously should be used for testing the extended gravity models though this method has its own shortcomings. If the orders of the post-Newtonian corrections of the extended gravity models lie beyond the PPN order it is natural to suggest that the parameters of the models should be limited via the second or the third post-Newtonian orders. The corresponding formalisms of the 2PN and 3PN really do exist \cite{damour, blanchet}. These formalisms consider the gravitational radiation and its subtle effects on pulsar timing and orbit parameters. However many calculations there are based on GR and do not suit for comparing arbitrary extended theories of gravity like the PPN formalism \cite{will} does.

There are also the other ways to test astrophysical predictions of the extended theories of gravity such as accretion onto massive objects and microlensing. After computing the accretion rate for some solution the result can be compared with GR predictions and some other extended gravity cases. The investigation of the data of gravity lensing events also is a perspective method as these data become more and more complete. Verification of the extended gravity models via studying binary systems and particularly the pulsar data needs special methods and approaches. Their construction is the subject of further considerations.

\section*{Acknowledgments}

The work was supported by Federal Agency on Science and Innovations of Russian Federation, state contract 02.740.11.0575. S.A. and B.L. also were supported by individual grants from Dmitry Zimin Foundation ``Dynasty''. Authors would like to thank Prof. S. Capozziello, Dr. M. Smolyakov and Dr. D. Levkov for useful discussions on the subject of this work. This work was performed at the Jet Propulsion Laboratory, California Institute of Technology, under a contract with the National Aeronautics and Space Administration.

\end{document}